# Soft Computing Framework for Routing in Wireless Mesh Networks: An Integrated Cost Function Approach

Shakti Kumar[1], Brahmjit Singh[2], Sharad Sharma[2]
[1]*Computational Intelligence (CI) Lab, IST Klawad, Yamunanagar, India*
[2]*Deptt. of Electronics & Comm. Engg., National Institute of Technology, Kurukshetra, India*
Email: shaktik@gmail.com , brahmjit.s@gmail.com, sharadpr123@rediffmail.com

*Abstract:* Dynamic behaviour of a WMN imposes stringent constraints on the routing policy of the network. In the shortest path based routing the shortest paths needs to be evaluated within a given time frame allowed by the WMN dynamics. The exact reasoning based shortest path evaluation methods usually fail to meet this rigid requirement. Thus, requiring some soft computing based approaches which can replace "best for sure" solutions with "good enough" solutions. This paper proposes a framework for optimal routing in the WMNs; where we investigate the suitability of Big Bang-Big Crunch (BB-BC), a soft computing based approach to evaluate shortest/near-shortest path. In order to make routing optimal we first propose to replace distance between the adjacent nodes with an integrated cost measure that takes into account throughput, delay, jitter and residual energy of a node. A fuzzy logic based inference mechanism evaluates this cost measure at each node. Using this distance measure we apply BB-BC optimization algorithm to evaluate shortest/near shortest path to update the routing tables periodically as dictated by network requirements. A large number of simulations were conducted and it has been observed that BB-BC algorithm appears to be a high potential candidate suitable for routing in WMNs.

I. INTRODUCTION

Wireless Mesh Networks are rapidly deployable, dynamically self organizing, self configuring, self balancing and self healing dynamic networks. In WMNs, each node (stationary or mobile) has the capability to join and create a network automatically by sensing other nodes within its coverage area. In WMNs, data hops from one device to another until it reaches a given destination and the network is established using multi-hop phenomena. Each node may act as a host as well as a router. WMN nodes are of two types: (1) Mesh Routers and (2) Mesh Clients. With the use of wireless network interface cards (NICs) conventional nodes (e.g., phones, desktops, laptops, PDAs, etc.) can connect directly to wireless mesh routers to establish a WMN. Minimal upfront investments are required to deploy WMNs as it can be deployed incrementally as needed, one node at a time. WMNs can be categorized to three types: (1) Infrastructure Mesh; (2) Client Mesh and (3) Hybrid Mesh. Infrastructure type WMNs are generally comprised of mesh routers forming an infrastructure for client nodes. Client mesh just like Mobile Ad-hoc Networks (MANETs) provides peer-to-peer networks and client nodes must perform routing and self-configuration functions too. Hybrid Mesh is the combination of infrastructure and client meshing. Client nodes can access the network through mesh routers as well as directly meshing with other client nodes thus making it a practical WMN [1].

Routing is the process of directing data packets towards a given destination through multiple hops. Due to the dynamic network conditions of a WMN, routing requires significant attention and it must work in a decentralized self-configuring manner. Mobile Ad hoc Networks (MANETs) and WMNs share many common features. Thus, the routing protocols developed for MANETs can usually be applied to WMNs as well e.g. Dynamic Source Routing (DSR) [2], Ad hoc On-demand Distance Vector (AODV) [3], Topology Broadcast based Reverse Path Forwarding (TBRPF) [4] etc. have been used in both. Enough scalability is not provided by the present MAC and routing protocols applied to WMNs [5]. As the number of nodes or hops in a WMN increases, the throughput decreases critically. However, to design an efficient WMN, all existing protocols from the application layer to transport, network MAC, and physical layers need to be re-designed. The factors like resource allocation, interference avoidance and rate adaptation across multiple hops critically affects the performance of a WMN [6]. The various performance parameters used in a WMN can be categorized as per flow (delay, packet loss ratio, delay jitter, hop count, throughput and interference); per node (computation complexity and power efficiency); per link (link quality, channel utilization, transmission rate, and congestion); inter flow (interference and fairness) and network wide parameters (QoS, total throughput etc) [1]. It is very difficult to invent a routing metric to include all these parameters due to being inter-related. Routing metrics developed so far take account of a subset of these performance parameters only. These routing metrics are Airtime Cost Routing Metric [6], Hop Count, Per-Hop Round Trip Time (RTT) [7], Metric of Interference and Channel-Switching (MIC) [8], Expected Transmission Count (ETX) [9], Expected Transmission on a Path (ETOP) [10], Expected Transmission Time (ETT) and Weighted Cumulative ETT (WCETT) [11], Low Overhead Routing Metric [12], Effective Number of Transmissions (ENT) [13], Per-Hop Packet Pair Delay and Expected Data Rate (EDR) [14], etc. The simplest one is Hop Count Metric however; in most cases the minimum hop-count is not sufficient for a routing protocol due to dynamic traffic conditions and propagation delays. Other routing metrics are severely affected with high complexity, in-appropriate load balancing, high overhead or low performance. The comparison of performance metrics on a routing algorithm is discussed in [14].

Due to complexities associated with exact reasoning there is an increasing demand for soft computing based techniques in WMN research. These techniques may make the WMNs more popular in terms of its self organizing and self configuring capabilities. Soft computing provides the optimal solution within an affordable time permitted by the WMN dynamics. Since, computation times are very short, the best solution needs to be replaced by good enough solutions such that a given WMN quickly adapts to dynamic environmental changes [15]. The Big Bang Big Crunch (BB-BC) optimization is one such approach. In this paper we propose a frame work in which we apply shortest path routing. The BB-BC approach finds the shortest path based upon an integrated distance measure consisting of throughput, delay, jitter and residual energy of a node. Fuzzy approach has been used to evaluate this integrated measure. The selected near shortest path given by BB-BC can be used for appropriate route selection in WMNs.

This paper is divided into six sections. Section I presents the motivation for the present work. Section II presents the system model, WMN node architecture and integrated cost function based frame work. Section III presents integrated cost evaluation method. The method is based upon fuzzy logic based approach. The integrated cost measure is the function of throughput, delay, jitter and residual energy of the WMN nodes. Section IV introduces the Big Bang Big Crunch (BB-BC) algorithm and its application to shortest/near shortest path evaluation amongst the many alternatives. The section V highlights and discusses the performance of BB-BC algorithm. Section VI concludes the paper.

## II. SYSTEM MODEL

Figure 1 shows the proposed architecture for network nodes. Each node of WMN can also act as a router and forward the data packets to other nodes. There may be many inputs (arriving from adjacent nodes Ni1 to Nin) and many output (towards Node No1 to Nom) nodes. The node processing system (NPS) decides which link is best for data packets to be transmitted and uses a queuing system to assign time slots for these data packets. The link state information is provided from this node processing system to a module which extracts the information of various link parameters e.g. throughput, end to end delay, jitter etc. of various links and residual energy of a node. Based on this information of these parameters a cost function based routing metric has been devised to find the cost of each link. This routing information is updated periodically or can be governed by the occurrence of some event. This new information updates the existing information in the Node Processing System for routing the data packets.

In order to analyse and optimize the performance of routing algorithm of WMN simulations were performed for a dynamic scenario in MATLAB. In this simulation, networks of 25, 50, 100, 500 and 1000 nodes are considered that are placed within a 500m X 500m and 1500m X 1500m area. The transmission range of the nodes is 250 meters. In all the network models node 1 acts as source and transmits data packets to the last node which is the terminal node. This communication is possible through hoping via various neighbouring nodes. In this type of wireless communication multiple routes/paths are available. Decision as regard to which path or route is to be used for any type of traffic, depends upon the current value of the integrated cost measure (distance) of the link connecting two adjacent nodes.

The system level algorithm for the frame work is summarized as follows:
Create a Wireless Mesh Network of specific set of connected nodes.
Display this network showing the architecture of WMN.
Using fuzzy logic approach evaluates the integrated cost of each link using delay, throughput, jitter and residual energy of each node.
For each source and terminal node pair using BB-BC algorithm evaluate the shortest/near shortest path between source to terminal node.
Update routing tables.
Wait for new network configuration to emerge/wait for permissible time interval.
Go to step 1.
A simple identification software takes care of updating the incoming and out-going nodes. Thus a WMN always remains updated. Cost function (distance between the two adjacent nodes) deciding the shortest path is a critical parameter. Section three below presents the integrated cost evaluation approach.

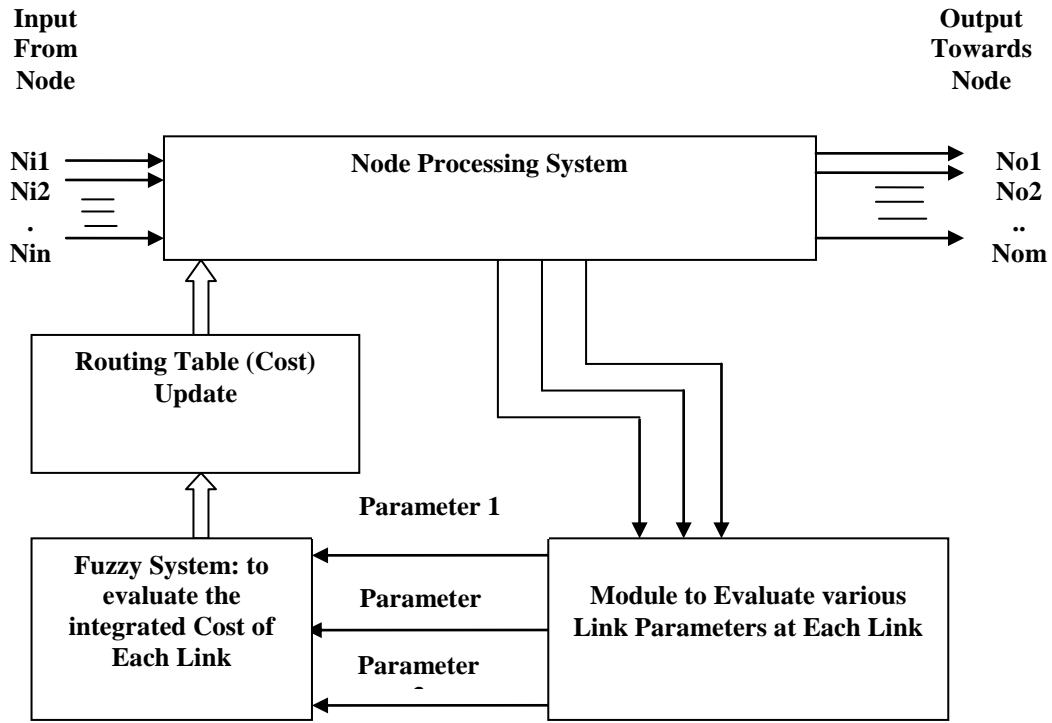

Figure 1. Node Architecture

III. Evaluating the Integrated cost Measure

It is a proven fact now that Fuzzy Logic is a powerful problem-solving methodology with wide range of applications in industrial control, consumer electronics, management, medicine, expert systems and information technology. It finds its applications for embedded system design as well. It presents a simple way to draw definite conclusions from vague, ambiguous or imprecise and incomplete information. It is very close to the way the human beings think and make decisions even under highly dynamic uncertain environments. Figure 2 shows a basic system consisting of four major modules of the system: fuzzification, inference engine, knowledge base and defuzzification module [16].

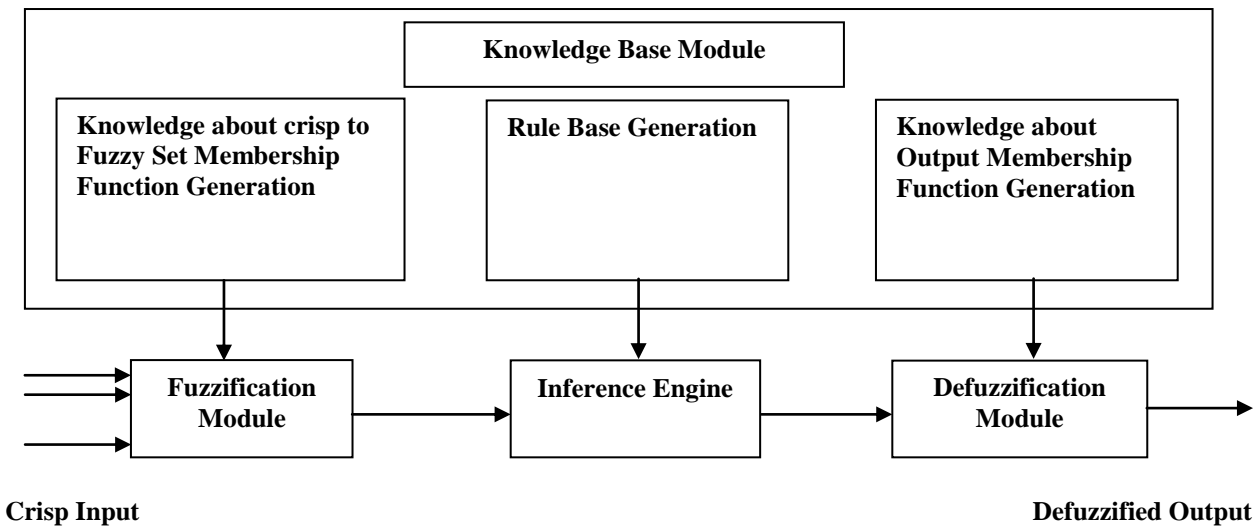

Figure 2. A Basic Fuzzy System

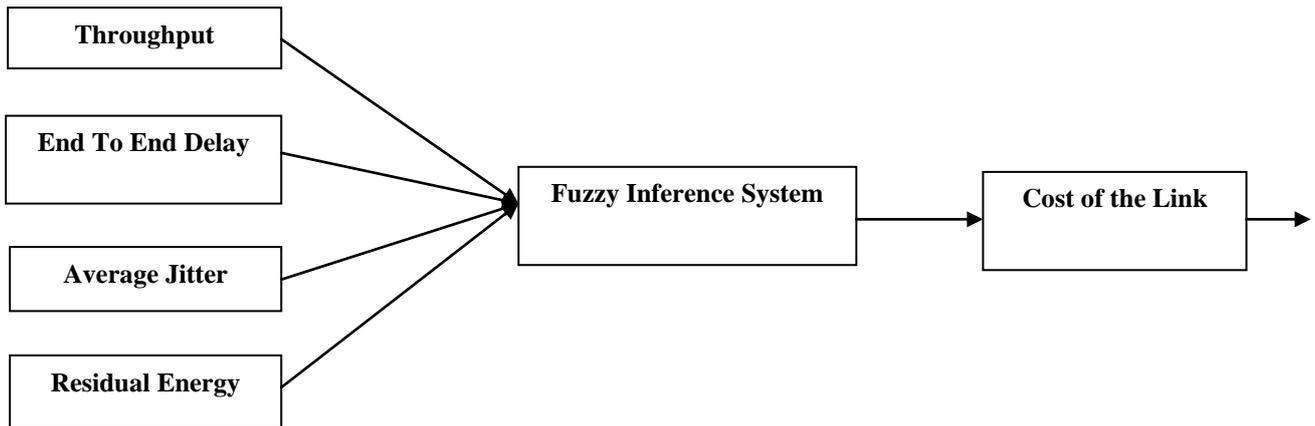

Figure 3: Cost Function generated using Fuzzy Logic

The fuzzification module transforms the crisp input(s) into fuzzy values. These values are then processed in fuzzy domain by inference engine based on the knowledge base (rule base and procedural knowledge) supplied by the domain expert(s). Finally the processed output is transformed from fuzzy domain to crisp domain by defuzzification module.

Figure 3 shows the integrated cost function measure proposed in this work. It consists of four significant parameters of the network: throughput, end-to-end delay, jitter of the link and the residual energy of the node. For a link high throughput, low end-to-end delay and low jitter are the suitable conditions. Here residual energy of a node is also added as in WMNs due to multipath hoping various paths are available and if node is positioned at a significant place then its energy will also play an important role. The node having less energy must transmit less energy and select the near most neighbours only. Based upon these four parameters the evaluated integrated cost measure is used as the distance between the two particular adjacent links.

## IV. BIG BANG BIG CRUNCH (BB-BC) ALGORITHM

The Big Bang theory is one of the most widely accepted theories explaining the evolution of our universe [17]. According to this theory, our universe suddenly sprang into existence as around 14 billion years ago. Our universe is thought to have inflated from an infinitesimally small, infinitely hot, infinitely intense pressure point. The pressure is thought to be so intense that finite matter is actually squished into infinite density. It inflated into existence in a split second giving it the name "Big Bang". There was nothing before this big bang and everything started evolving after the bang; the planets, the galaxies etc. The energy generated by the Big Bang i.e. kinetic energy is counterbalanced by the gravitational pull between these entities. If this gravitational energy becomes greater than the kinetic energy, the expansion of the universe stops and the universe will start to contract which will reduce it to a single point with both infinite density and temperature. This contraction phase is called Big Crunch Phase. This happens every about 14 billion years.
The Big Bang Big Crunch (BB-BC) optimization was introduced by Erol and Eksin [17]. The optimization method is inspired by the Big Bang and Big Crunch theory of evolution of the universe. The method has been well received by the research community as it has a low computational complexity and high convergence speed. BB-BC algorithm considers that in the Big Bang phase energy dissipation produces disorder and randomness; whereas, in the Big Crunch phase, randomly distributed particles are drawn into an order. The BB-BC algorithm produces a series of sequential Big Bangs and Big Crunches. The distribution of randomness within the search space produced during the Big Bang phase becomes smaller and smaller about the average point computed during the Big Crunch. After some iterations the algorithm converges to a solution. The BB–BC method has been shown to outperform the enhanced classical Genetic Algorithm for many benchmark test functions [17].

### A. BB-BC Algorithm:

The BB-BC algorithm [18] is summarized as given below:
Notations:
$G_0 (V_0, E_0)$ initial WMN topology graph;
$G_i (V_i, E_i)$ WMN topology graph after the $i$th change;
$s$ source node;
$t$ destination node;
$P_i(s, t)$ path from $s$ to $t$ on the graph $G_i$;
$d_l$ transmission delay on the communication link $l$;
$c_l$ Integrated cost on the communication link $l$;
$\Delta (P_i)$ total transmission delay on the path $P_i$;
$C (P_i)$ the total integrated cost of the path $P_i$.

### B. BB-BC Algorithm for DSPRP

Step1:

Form an initial generation of N candidates in a random manner as given below:
We start to search a random path from $s$ to t by randomly selecting a node $v_1$ from $N(s)$, the neighbourhood of $s$. Then, we randomly select a node $v_2$ from $N (v_1)$. We continue with this process until $t$ is reached. In order to keep the path loop-free, those nodes that are already included in the current path are excluded from being selected as the next node to be added into the path, thereby avoiding re-entry of the same node into a path. This gives us a random path $P(s, t) = \{s, v_1, v_2, \ldots t\}$. Repeating this process for $N$ times, we get the initial population

$N = \{Ind_0, Ind_1, \ldots, Ind_{q-1},\}$ .

Respect the limits of the search space.

Step2:
Using a fuzzy system evaluate the integrated cost (fitness) of each path.

Step3:
Find the center of mass according to equation (1).

$$x_c = \frac{\sum_{i=1}^{N} \frac{1}{f^i} x_i}{\sum_{i=1}^{N} \frac{1}{f^i}} \quad (1)$$

where $x_c$ = position of the centre of mass; $x_i$ = position of candidate; $f^i$ = fitness function value of candidate i; $N$ = population size.

(Alternatively best fitness individual can be chosen as the centre of mass instead of using (1)).

Step4:
Generate new population (set of paths) with some randomness from the fittest candidate.

Step5:
if stopping criteria has not been met then go to Step 2. Otherwise display the shortest (near shortest) path enumerated.
stop

The algorithm was applied to the proposed framework to evaluate the optimal shortest path under given time constraints. The results are discussed in section V for various time and network configurations.

## V. RESULTS AND DISCUSSION

The BB-BC algorithm was implemented in MATLAB. Numerical results were computed and are given in Table 1. It has been observed from the results of Table 1 that this frame work has a practical approach. This frame work respects the dynamic architecture of WMN with varying number of nodes and iterations. It is observed that the path cost of the shortest path obtained with various WMN models is quite same as the practical network. It is worth mentioning here that this frame work satisfies the random and dynamic behavior of a WMN as with 1000 nodes and 100 iterations it is finding the shortest path with the minimum path cost amongst all network models. Results show that when network size exceeds 100 nodes the time taken by BB-BC increases significantly although providing the shortest path accurately. When WMN size is of the order of 1000 nodes it is still getting a direct path between source and transmitter due to the dynamic and random behavior of a WMN. As the network size is 1000 nodes the time taken by BB-BC to find the near shortest path is 50 minutes with 200 iterations proving that the network size is a limiting factor for optimal performance of a WMN. But as WMN complexity grows we may not get the shortest path in the time frame allowed for this process. Under such condition we have to settle for the good enough and not the best concept. Figures 4 and 6 show the orientation of WMN for 25 nodes and 100 and 200 iterations respectively and figures 8 and 10 show the orientation of WMN for 50 nodes and 100 and 200 iterations respectively. As the size of the WMN increases the plot is not much visible due to various links hence, it is not presented in this paper.

As many nodes shall be joining the network and many shall be leaving network, the number of iterations allowed for evaluating shortest/near shortest path will depend upon the time allowed between the subsequent WMN updates.

Table 1: Results of BB-BC algorithm with various network size

| No. Of Nodes | No. Of Generations | Path Cost | Time (Sec) | Shortest Path |
|---|---|---|---|---|
| 25 | 100 | 0.2653 | 2.214822 | 1-19-22-25 |
| 25 | 200 | 0.1390 | 2.940108 | 1-6-25 |
| 50 | 100 | 0.1889 | 3.398222 | 1-30-46-50 |
| 50 | 200 | 0.1274 | 6.284940 | 1-16-50 |
| 100 | 100 | 0.1795 | 10.542362 | 1-8-100 |
| 100 | 200 | 0.4044 | 15.953822 | 1-14-64-29-100 |
| 500 | 100 | 0.7158 | 196.735260 | 1-87-404-218-389-240-315-500 |
| 500 | 200 | 0.7144 | 417.595691 | 1-491-140-180-354-307-116-33-500 |
| 1000 | 100 | 0.1126 | 1282.580075 | 1-1000 |
| 1000 | 200 | 0.8140 | 2982.169662 | 1-864-706-45-416-13-767-308-1000 |

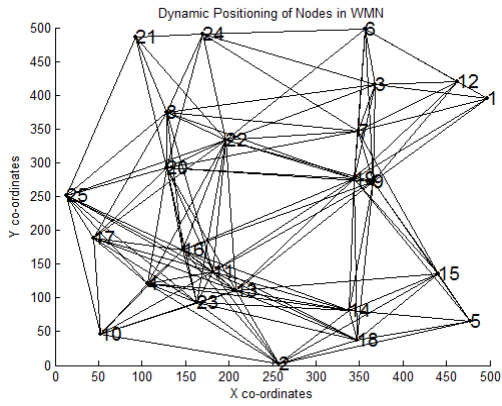

Figure 4: Plot of WMN for 25 Nodes, 100 Iterations

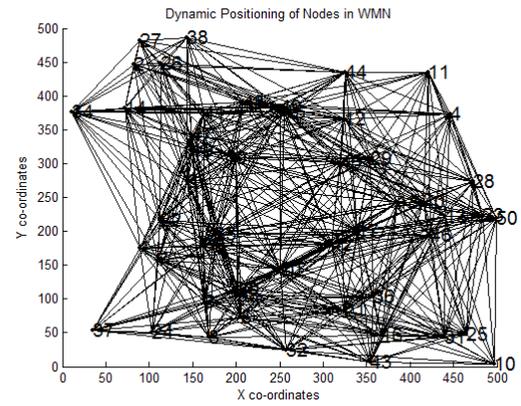

Figure 8: Plot of dynamic WMN for 50 Nodes, 100 Iterations

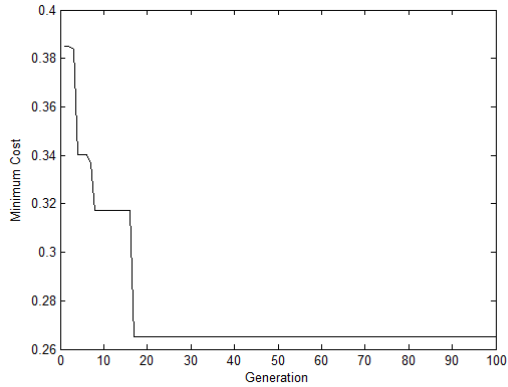

Figure 5: Plot of Path Cost v/s Generation for 25 Nodes, 100 Iterations

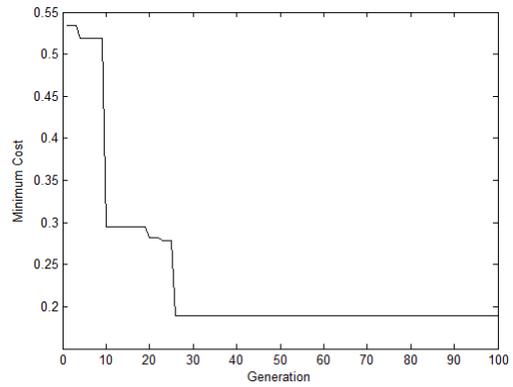

Figure 9: Plot of Path Cost v/s Generation for 50 Nodes, 100 Iterations

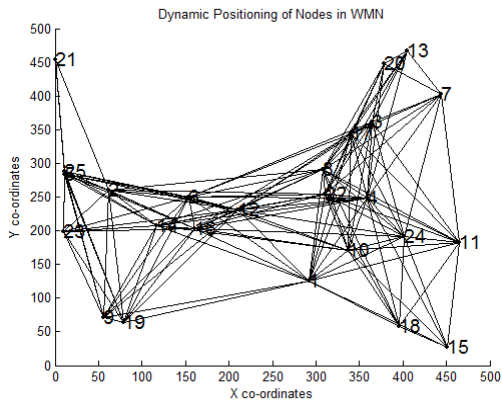

Figure 6: Plot of dynamic WMN for 25 Nodes, 200 Iterations

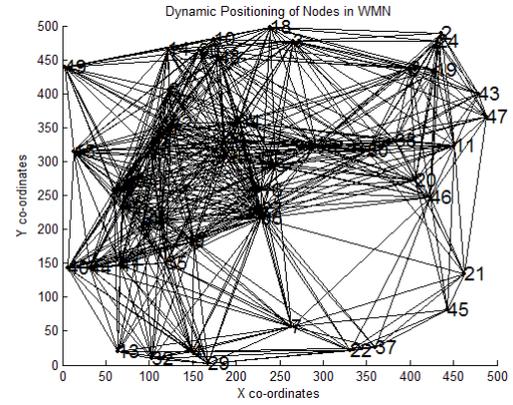

Figure 10: WMN for 50 Nodes, 200 Iterations

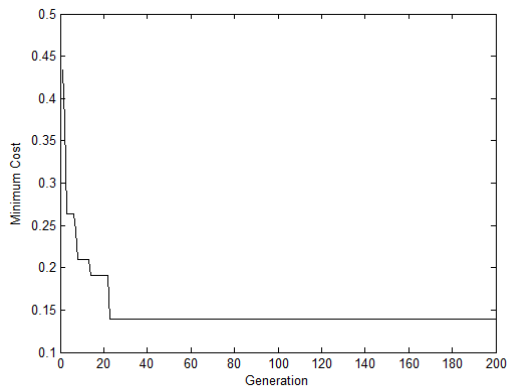

Figure 7: Plot of Path Cost v/s Generation for 25 Nodes, 200 Iterations

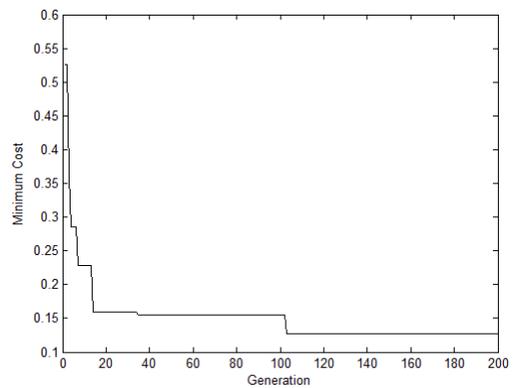

Figure 11: Plot of Path Cost v/s Generation for 50 Nodes, 200 Iterations

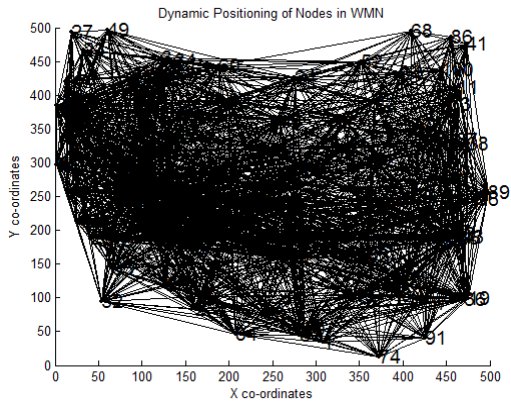

Figure 12: Plot of WMN for 100 Nodes, 100 Iterations

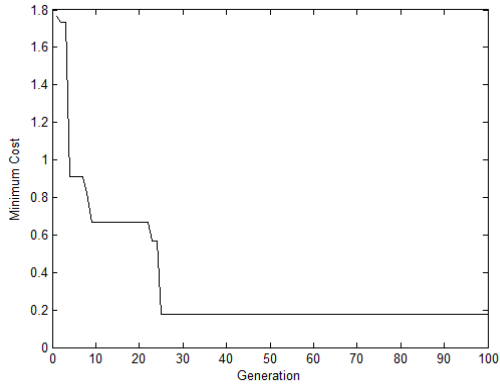

Figure 13: Plot of Path Cost v/s Generation for 100 Nodes, 100 Iterations

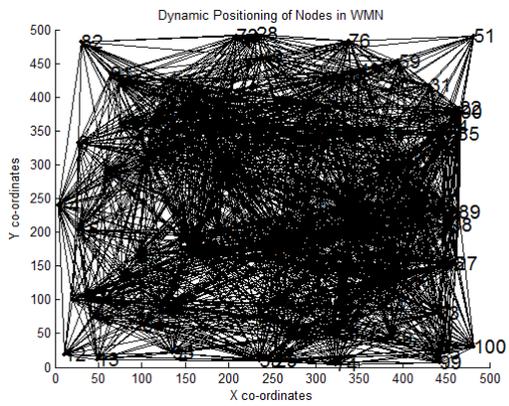

Figure 14: Plot of WMN for 100 Nodes, 200 Iterations

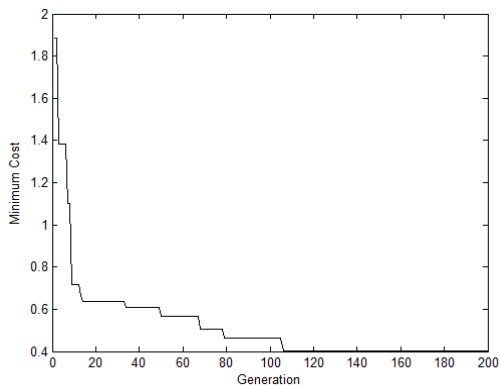

Figure 15: Plot of Path Cost v/s Generation for 100 Nodes, 200 Iterations

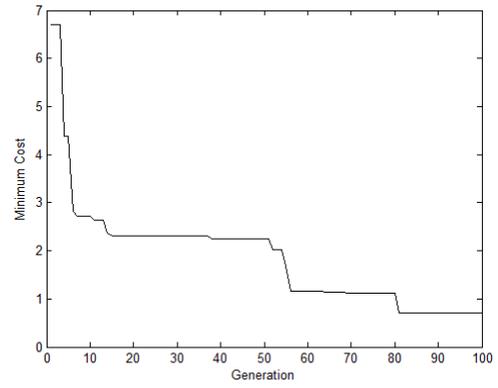

Figure 16: Plot of Path Cost v/s Generation for 500 Nodes, 100 Iterations

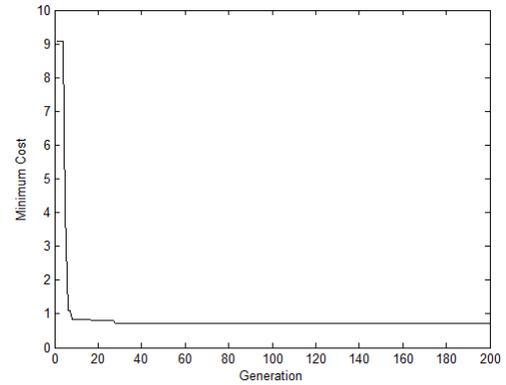

Figure 17: Plot of Path Cost v/s Generation for 500 Nodes, 200 Iterations

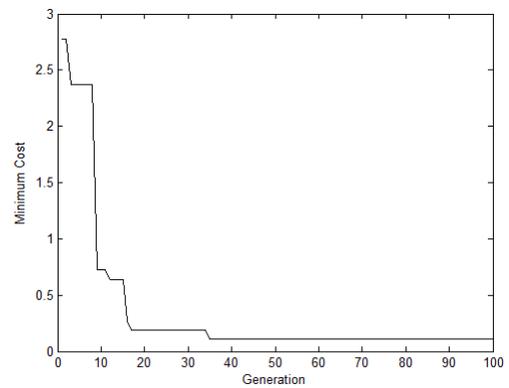

Figure 18: Plot of Path Cost v/s Generation for 1000 Nodes, 100 Iterations

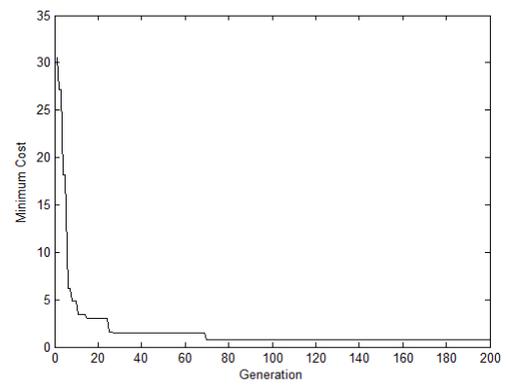

Figure 19: Plot of Path Cost v/s Generation for 1000 Nodes, 200 Iterations

## VI. Conclusion

WMNs are dynamic, self organizing, self configuring and self balancing wireless networks. This paper proposes an integrated cost function based framework for routing in WMNs and investigates the suitability of BB-BC algorithm to routing in WMNs. Due to its highly dynamic behaviour, WMNs must be capable of evaluating the shortest path or a near shortest path as quickly as possible. As the complexity and size of the network grows, time requirement to find the shortest path is to be compromised. In order to evaluate the application of of BB-BC algorithm to WMNs we first enumerated an integrated cost measure consisting of throughput, delay, jitter and residual energy of the nodes. Using this distance measure between the adjacent node pairs we performed a large number of simulations to find the shortest path in a 25, 50, 100, 500 and 1000 node WMNs. It was observed that BB-BC successfully converged to an optimal solution in very less time. Time limit can be stipulated based upon the network size and the trade off between accuracy and allowable time has to be accepted for optimal routing performance. It is also established that as the network size increases, the evaluation time to find the shortest path also increases significantly thus requires some cluster heads to provide the information locally in the micro cell area or subnets. From our observations we conclude that soft computing approaches to shortest path routing can appropriately handle the constraints imposed by network dynamics of WMNs. BB-BC is a much suited potential candidate for routing application in a WMN.

## References

[1] Akyildiz IF, Wang X and Wang W "Wireless mesh networks: a survey", Computer Networks Journal (Elsevier) 47(4), 2005, 445–487.
[2] D.B.Johnson, D.A.Maltz and Y.C.Hu, "The dynamic source routing protocol for mobile ad hoc networks (DSR)", IETF, July 2004.
[3] C.Perkins, E.Belding-Royer and S.Das, "Ad hoc on-demand distance vector (AODV) routing", IETF RFC 3561, July 2003.
[4] Ogier R, Templin F and LewisM, "Topology dissemination based on reverse-path forwarding (TBRPF)", IETF RFC 3684, 2004
[5] L. Huang, T. Lai, "On the scalability of IEEE 802.11 ad hoc networks", ACM International Symposium on Mobile Ad Hoc Networking and Computing (MOBIHOC), 2002, pp. 173–182.
[6] Akyildiz IF, Wang X and Wang W "Wireless mesh networks", Wiley, 2009.
[7] A.Adya, P.Bahl, J.Padhye, A.Wolman and L.Zhou, "A multi radio unification protocol for IEEE 802.11 wireless networks", International Conference on Broadcast Networks (Broad Nets), 2004, pp.344- 354.
[8] Yang Y, Wang J and Kravets, "Interference-aware load balancing for multihop wireless networks", In Tech. Rep. UIUCDCS-R-2005-2526, Department of Computer Science, University of Illinois at Urbana-Champaign, 2005.
[9] De Couto DSJ, Aguayo D, Bicket J and Morris R, "A high-throughput path metric for multihop wireless routing", In Proc. ACM Annual International Conference on Mobile Computing and Networking (MOBICOM), 2003, pp. 134–146.
[10] Jakllari G, Eidenbenz S, Hengartner N, Krishnamurthy S and Faloutsos M, "Link positions matter: a noncommutative routing metric for wireless mesh networks", In Proc. IEEE Annual Conference on Computer Communications (INFOCOM) 2008, pp. 744-752.
[11] R.Draves, J.Padhye and B.Zill, "Routing in Multi Radio, Multi-Hop Wireless Mesh Networks," ACM Annual International Conference on Mobile Computing and Networking (MobiCom'04), pp. 114-128.
[12] Karbaschi G and Fladenmuller A, "A link quality and congestion-aware cross layer metric for multi-hop wireless routing", In Proc. of IEEE MASS'05, pp. 7–11.
[13] Koksal CE and Balakrishnan H, "Quality-aware routing metrics for time-varying wireless mesh networks" IEEE Journal on Selected Areas in Communications 24(11), 2006 pp. 1984–1994.
[14] R. Draves, J. Padhye, B. Zill, "Comparisons of routing metrics for static multi-hop wireless networks", ACM Annual Conference of the Special Interest Group on Data Communication (SIGCOMM), August 2004, pp. 133–144.
[15] Shengxiang Yang, Member, IEEE, Hui Cheng, and Fang Wang, "Genetic Algorithms With Immigrants and Memory Schemes for Dynamic Shortest Path Routing Problems in Mobile Ad Hoc Networks", IEEE Transactions on Systems, MAN, and Cybernetics—Part C: Applications and Reviews, vol. 40, No. 1, January 2010, pp. 52-63.
[16] John Yen and Reza Langari, "Fuzzy Logic Intelligence, Control and Information," Prentice Hall, New Jersey, 1999.
[17] O.K.Erol and I.Eksin, "A new optimization method: Big Bang-Big Crunch" Advances in Engineering Software, 37, 2006, pp. 106-111.
[18] Y. Labbi, D. Ben Attous, "Big Bang–Big Crunch Optimization Algorithm for Economic Dispatch with Valve-Point Effect", Journal of Theoretical and Applied Information Technology, 2010, pp. 48-56.